\documentclass[showpacs,aps,prd]{revtex4}
\input epsf

\begin{document}

\title{$Y(4626)$ as a $P$-wave $[cs][\bar{c}\bar{s}]$ tetraquark state}
\author{Jian-Rong Zhang}
\affiliation{Department of Physics, College of Liberal Arts and Sciences, National University of Defense Technology,
Changsha 410073, Hunan, People's Republic of China}


\begin{abstract}
Motivated by the Belle Collaboration's new observation of $Y(4626)$,
we investigate the possibility of its configuration
as a $P$-wave $[cs][\bar{c}\bar{s}]$ tetraquark state
from QCD sum rules. Eventually, the
extracted mass $4.60^{+0.13}_{-0.19}~\mbox{GeV}$ for the
$P$-wave
$cs$-scalar-diquark $\bar{c}\bar{s}$-scalar-antidiquark state
agrees well with the
experimental data of $Y(4626)$,
which could support its interpretation as a $P$-wave
scalar-scalar $[cs][\bar{c}\bar{s}]$ tetraquark state.
\end{abstract}
\pacs {11.55.Hx, 12.38.Lg, 12.39.Mk}\maketitle

\section{Introduction}\label{sec1}
Very newly, Belle Collaboration reported the first observation of a vector
charmoniumlike state $Y(4626)$ decaying to a charmed-antistrange
and anticharmed-strange meson pair
$D_{s}^{+}D_{s1}(2536)^{-}$
with a significance of $5.9\sigma$ \cite{Y4626}. Its mass
and width were measured to be $4625.9_{-6.0}^{+6.2}
\pm0.4~\mbox{MeV}$
and $49.8_{-11.5}^{+13.9}
\pm4.0~\mbox{MeV}$, respectively.
This state is near the $Y(4660)$
observed in the hidden-charm
process $e^{+}e^{-}\rightarrow\psi(2S)\pi^{+}\pi^{-}$ \cite{Y4660,Y4660-1}
and also consistent with the $Y(4630)$ searched in the
$e^{+}e^{-}\rightarrow\Lambda_{c}\bar{\Lambda}_{c}$ \cite{Y4630,Y4630-2}.
Considering their close masses and widths,
$Y(4660)$ and $Y(4630)$ were
suggested to be the same resonance \cite{Y4630-Y4660,Y4630-Y4660-1,Y4630-Y4660-2},
and there have been various theoretical explanations for them,
such as a conventional charmonium \cite{Charmonium,Charmonium-1,Charmonium-2},
a $f_{0}(980)\Psi'$ bound state \cite{molecular,molecular-1,molecular-2}, a baryonium
state \cite{Y4630-Y4660,baryonium,baryonium-2}, a hadro-charmonium state \cite{Hadro-Charmonium}, a
tetraquark state \cite{tetraquark,tetraquark-1,Y4660-QCDSR,Y4660-QCDSR-1,Y4660-QCDSR-2,Y4660-QCDSR-3,Y4660-QCDSR-4} and so on.

The new observation of $Y(4626)$ by Belle immediately aroused one's great
interest \cite{Th-charmonium,Th-molecular,Th-molecular-1,Th-molecular-2,Th-tetraquark,Th-tetraquark-1,Th-tetraquark-2}.
With an eye to the multiquark viewpoint, an assignment
of $Y(4626)$ was proposed as a $D_{s}^{*}\bar{D}_{s1}(2536)$ molecular state
in a quasipotential Bethe-Salpeter equation
approach with
the one-boson-exchange model \cite{Th-molecular}. Later,
the mass spectrum
of $D_{s}^{*}\bar{D}_{s1}(2536)$ system
was calculated within the framework
of Bethe-Salpeter equations \cite{Th-molecular-2}, and in the end
the authors may not think $Y(4626)$ to be a $D_{s}^{*}\bar{D}_{s1}(2536)$ bound state,
but something else. Otherwise,
some authors employed a multiquark
color flux-tube model with a multibody confinement potential and one-glue-exchange interaction
to make an exhaustive investigation on the diquark-antidiquark state \cite{Th-tetraquark-1}, and they
concluded that $Y(4626)$ can be well interpreted as
a $P$-wave $[cs][\bar{c}\bar{s}]$ state.

Under the circumstance, it is interesting and
significant to study that whether $Y(4626)$ could be a candidate of $P$-wave
$[cs][\bar{c}\bar{s}]$ tetraquark state by different means.
It is known that one has to
face the complicated nonperturbative problem in QCD
while handling a hadronic state.
Established firmly on the QCD basic theory,
the QCD sum rule \cite{svzsum} acts as one
authentic way for evaluating nonperturbative effects,
which has been successfully applied to
plenty of hadronic systems (for reviews see
\cite{overview,overview1,overview2,overview3} and references
therein).
Therefore, in this work we devote to investigating
that whether $Y(4626)$ could be a $P$-wave
$[cs][\bar{c}\bar{s}]$ tetraquark state with the
QCD sum rule method.

This paper is organized as follows. The QCD sum rule
for the $P$-wave tetraquark state is derived in Sec. \ref{sec2}, followed
by the numerical analysis in Sec.
\ref{sec3}. The last part is a brief summary and outlook.

\section{the $P$-wave $[cs][\bar{c}\bar{s}]$ state QCD sum rule}\label{sec2}
Generally, it is possible to get a total spin parity $1^{-}$ for a $P$-wave tetraquark with
several diquark choices.
Meanwhile, one could note that there have been broad discussions on the
``good" diquarks or ``bad"
diquarks in the tetraquark configurations \cite{diquarks}.
Thus, one could represent the $P$-wave $[cs][\bar{c}\bar{s}]$ tetraquark state
basically from following considerations \cite{current}.
A good diquark operator in the attractive anti-triplet color channel
can be written as $\bar{q}_{c}\gamma_{5}q$,
and a bad diquark operator can be written as $\bar{q}_{c}\gamma q$,
which represent $0^{+}$ and $1^{+}$, respectively.
Similarly, one can construct $0^{-}$ and $1^{-}$ operators as $\bar{q}_{c}q$
and $\bar{q}_{c}\gamma\gamma_{5}q$, respectively. Lattice studies suggest that diquarks are preferably (energetically) formed
into spin $0$ configurations \cite{Lattice}.
In fact, the solid tetraquark candidates tend to be made of good diquarks.
As the concrete examples of QCD sum rules, we have explored some tetraquark states with
various diquark configurations \cite{scalar-scalar},
and the final results favored the scalar diquark-scalar antidiquark structures,
which virtually manifests that a solid tetraquark state should be composed of good diquarks.

In this manner, the $P$-wave $[cs][\bar{c}\bar{s}]$ state could be described as
having the flavor content $[cs][\bar{c}\bar{s}]$ with the spin momentum
numbers $S_{[cs]}=0$, $S_{[\bar{c}\bar{s}]}=0$, and $S_{[cs][\bar{c}\bar{s}]}=0$, as well as
with the orbital momentum
number $L_{[cs][\bar{c}\bar{s}]}=1$.
To characterize such a state, one could first construct
the current
\begin{eqnarray}
j_{\mu}=\epsilon_{def}\epsilon_{d'e'f}(s_{d}^{T}C\gamma_{5}c_{e})D_{\mu}(\bar{s}_{d'}\gamma_{5}C\bar{c}_{e'}^{T}),
\end{eqnarray}
for the $P$-wave scalar-scalar case,
and then \begin{eqnarray}
j_{\mu}=\epsilon_{def}\epsilon_{d'e'f}(s_{d}^{T}Cc_{e})D_{\mu}(\bar{s}_{d'}C\bar{c}_{e'}^{T}),
\end{eqnarray}
for the $P$-wave pseudoscalar-pseudoscalar case.
Here the index $T$ means matrix
transposition, $C$ denotes the charge conjugation matrix,
the covariant derivative $D_{\mu}$ is introduced to generate $L=1$, and $d$, $e$,
$f$, $d'$, and $e'$ are color indices.

In general, the two-point correlator
\begin{eqnarray}
\Pi_{\mu\nu}(q^{2})=i\int
d^{4}x\mbox{e}^{iq.x}\langle0|T[j_{\mu}(x)j_{\nu}^{+}(0)]|0\rangle,
\end{eqnarray}
can be parameterized as
\begin{eqnarray}
\Pi_{\mu\nu}(q^{2})=\frac{q_{\mu}q_{\nu}}{q^{2}}\Pi^{(0)}(q^{2})+(\frac{q_{\mu}q_{\nu}}{q^{2}}-g_{\mu\nu})\Pi^{(1)}(q^{2}).
\end{eqnarray}
Furthermore, the part of correlator proportional to $-g_{\mu\nu}$ is
used to obtain the sum rule,
which can be evaluated in two different ways: at the hadronic
level and at the quark level.
Phenomenologically,
$\Pi^{(1)}(q^{2})$ can be written as
\begin{eqnarray}\label{Ph}
\Pi^{(1)}(q^{2})=\frac{\lambda^{2}}{M_{H}^{2}-q^{2}}+\frac{1}{\pi}\int_{s_{0}}
^{\infty}ds\frac{\mbox{Im}\Pi^{(1)}(s)}{s-q^{2}},
\end{eqnarray}
where $M_{H}$ denotes the hadron's mass. In the OPE
side, it can be expressed as
\begin{eqnarray}\label{OPE}
\Pi^{(1)}(q^{2})=\int_{(2m_{c}+2m_{s})^{2}}^{\infty}ds\frac{\rho(s)}{s-q^{2}},
\end{eqnarray}
for which the spectral density
$\rho(s)=\frac{1}{\pi}\mbox{Im}\Pi^{(1)}(s)$.

To derive $\rho(s)$, one works at leading order in $\alpha_{s}$
and includes condensates up to dimension $8$.
The strange quark is treated as a light one and the
diagrams are considered up to the order $m_{s}$.
Keeping the heavy-quark mass finite, one uses
the heavy-quark propagator in
momentum space \cite{reinders}.
The correlator's light-quark part
is calculated in the
coordinate space and Fourier-transformed to the momentum
space in $D$ dimension, which is combined
with the heavy-quark part and then dimensionally regularized at
$D=4$ \cite{overview3,Nielsen,Zhang}. At length,
it is given by
$\rho(s)=\rho^{\mbox{pert}}+\rho^{\langle\bar{s}s\rangle}+\rho^{\langle
g^{2}G^{2}\rangle}+\rho^{\langle
g\bar{s}\sigma\cdot G s\rangle}+\rho^{\langle\bar{s}s\rangle^{2}}+\rho^{\langle g^{3}G^{3}\rangle}+\rho^{\langle\bar{s}s\rangle\langle g^{2}G^{2}\rangle}+\rho^{\langle\bar{s}s\rangle\langle g\bar{s}\sigma\cdot G s\rangle}$,
concretely with
\begin{eqnarray}
\rho^{\mbox{pert}}&=&-\frac{1}{3\cdot5\cdot2^{11}\pi^{6}}\int_{\alpha_{min}}^{\alpha_{max}}\frac{d\alpha}{\alpha^{4}}\int_{\beta_{min}}^{1-\alpha}\frac{d\beta}{\beta^{4}}(1-\alpha-\beta)\kappa
[r-5m_{c}m_{s}(\alpha+\beta)]r^{4},\nonumber\\
\rho^{\langle\bar{s}s\rangle}&=&\frac{\langle\bar{s}s\rangle}{3\cdot2^{6}\pi^{4}}\Bigg\{\int_{\alpha_{min}}^{\alpha_{max}}\frac{d\alpha}{\alpha^{2}}\int_{\beta_{min}}^{1-\alpha}\frac{d\beta}{\beta^{2}}\Big\{[(2-\alpha-\beta)m_{c}+(1-\alpha-\beta)m_{s}]r\nonumber\\
&&{}-3(\alpha-\alpha^{2}+\beta-\beta^{2})m_{s}m_{c}^{2}\Big\}r^{2}-m_{s}\int_{\alpha_{min}}^{\alpha_{max}}\frac{d\alpha}{\alpha(1-\alpha)}[m_{c}^{2}-\alpha(1-\alpha)s]^{3}\Bigg\},\nonumber\\
\rho^{\langle g^{2}G^{2}\rangle}&=&-\frac{m_{c}\langle
g^{2}G^{2}\rangle}{3^{2}\cdot2^{12}\pi^{6}}\int_{\alpha_{min}}^{\alpha_{max}}\frac{d\alpha}{\alpha^{4}}\int_{\beta_{min}}^{1-\alpha}\frac{d\beta}{\beta^{4}}(1-\alpha-\beta)(\alpha^{3}+\beta^{3})\kappa
r[(m_{c}-3m_{s})r-2m_{s}m_{c}^{2}(\alpha+\beta)],\nonumber\\
\rho^{\langle g\bar{s}\sigma\cdot G s\rangle}&=&\frac{\langle
g\bar{s}\sigma\cdot G
s\rangle}{3\cdot2^{8}\pi^{4}}\Bigg\{\int_{\alpha_{min}}^{\alpha_{max}}\frac{d\alpha}{\alpha^{2}}\int_{\beta_{min}}^{1-\alpha}\frac{d\beta}{\beta^{2}}r\Big\{-3m_{c}(\alpha+\beta-4\alpha\beta)r
+m_{s}\alpha\beta[12m_{c}^{2}-7(\alpha+\beta)m_{c}^{2}-5\alpha\beta s]\Big\}\nonumber\\
&&{}+\int_{\alpha_{min}}^{\alpha_{max}}d\alpha[m_{c}^{2}-\alpha(1-\alpha)s]\Big\{\frac{3m_{c}}{\alpha(1-\alpha)}[m_{c}^{2}-\alpha(1-\alpha)s]+2m_{s}[5\alpha(1-\alpha)s-9m_{c}^{2}]\Big\}\Bigg\},\nonumber\\
\rho^{\langle\bar{s}s\rangle^{2}}&=&\frac{m_{c}\varrho\langle\bar{s}s\rangle^{2}}{3\cdot2^{4}\pi^{2}}\int_{\alpha_{min}}^{\alpha_{max}}d\alpha\Big\{-2m_{c}[m_{c}^{2}-\alpha(1-\alpha)s]+m_{s}[m_{c}^{2}-2\alpha(1-\alpha)s]\Big\},\nonumber\\
\rho^{\langle g^{3}G^{3}\rangle}&=&-\frac{\langle
g^{3}G^{3}\rangle}{3^{2}\cdot2^{14}\pi^{6}}\int_{\alpha_{min}}^{\alpha_{max}}\frac{d\alpha}{\alpha^{4}}\int_{\beta_{min}}^{1-\alpha}\frac{d\beta}{\beta^{4}}(1-\alpha-\beta)\kappa
\Big\{[(\alpha^{3}+\beta^{3})r+4(\alpha^{4}+\beta^{4})m_{c}^{2}\nonumber\\
&&{}-2m_{c}m_{s}(2\alpha^{2}+3\alpha\beta+2\beta^{2})(3\alpha^{2}-4\alpha\beta+3\beta^{2})]r-4m_{s}m_{c}^{3}(\alpha+\beta)(\alpha^{4}+\beta^{4})\Big\},\nonumber\\
\rho^{\langle\bar{s}s\rangle\langle g^{2}G^{2}\rangle}&=&\frac{m_{c}\langle\bar{s}s\rangle\langle
g^{2}G^{2}\rangle}{3^{2}\cdot2^{8}\pi^{4}}\Bigg\{\int_{\alpha_{min}}^{\alpha_{max}}\frac{d\alpha}{\alpha^{2}}\int_{\beta_{min}}^{1-\alpha}\frac{d\beta}{\beta^{2}}\Big\{(2-\alpha-\beta)(\alpha^{3}+\beta^{3})m_{c}^{2}-3[\alpha^{2}(\beta-1)+\beta^{2}(\alpha-1)]r\nonumber\\
&+&[2(\alpha^{2}+\beta^{2})^{2}-(\alpha+\beta)^{3}-\alpha\beta(\alpha-\beta)^{2}-(\alpha^{3}+\beta^{3})]m_{s}m_{c}\Big\}
- m_{s}m_{c}\int_{\alpha_{min}}^{\alpha_{max}}d\alpha\frac{3\alpha^{2}-3\alpha+1}{\alpha(1-\alpha)}\Bigg\}
,\nonumber\\
\rho^{\langle\bar{s}s\rangle\langle g\bar{s}\sigma\cdot G s\rangle}&=&\frac{m_{c}(m_{c}-m_{s})\langle\bar{s}s\rangle\langle g\bar{s}\sigma\cdot G s\rangle}{3\cdot2^{5}\pi^{2}}\int_{\alpha_{min}}^{\alpha_{max}}d\alpha(6\alpha^{2}-6\alpha+1)
\nonumber
\end{eqnarray}
for the $P$-wave scalar-scalar case, and with
\begin{eqnarray}
\rho^{\mbox{pert}}&=&-\frac{1}{3\cdot5\cdot2^{11}\pi^{6}}\int_{\alpha_{min}}^{\alpha_{max}}\frac{d\alpha}{\alpha^{4}}\int_{\beta_{min}}^{1-\alpha}\frac{d\beta}{\beta^{4}}(1-\alpha-\beta)\kappa
[r+5m_{c}m_{s}(\alpha+\beta)]r^{4},\nonumber\\
\rho^{\langle\bar{s}s\rangle}&=&\frac{\langle\bar{s}s\rangle}{3\cdot2^{6}\pi^{4}}\Bigg\{\int_{\alpha_{min}}^{\alpha_{max}}\frac{d\alpha}{\alpha^{2}}\int_{\beta_{min}}^{1-\alpha}\frac{d\beta}{\beta^{2}}\Big\{[-(2-\alpha-\beta)m_{c}+(1-\alpha-\beta)m_{s}]r\nonumber\\
&&{}-3(\alpha-\alpha^{2}+\beta-\beta^{2})m_{s}m_{c}^{2}\Big\}r^{2}-m_{s}\int_{\alpha_{min}}^{\alpha_{max}}\frac{d\alpha}{\alpha(1-\alpha)}[m_{c}^{2}-\alpha(1-\alpha)s]^{3}\Bigg\},\nonumber\\
\rho^{\langle g^{2}G^{2}\rangle}&=&-\frac{m_{c}\langle
g^{2}G^{2}\rangle}{3^{2}\cdot2^{12}\pi^{6}}\int_{\alpha_{min}}^{\alpha_{max}}\frac{d\alpha}{\alpha^{4}}\int_{\beta_{min}}^{1-\alpha}\frac{d\beta}{\beta^{4}}(1-\alpha-\beta)(\alpha^{3}+\beta^{3})\kappa
r[(m_{c}+3m_{s})r+2m_{s}m_{c}^{2}(\alpha+\beta)],\nonumber\\
\rho^{\langle g\bar{s}\sigma\cdot G s\rangle}&=&\frac{\langle
g\bar{s}\sigma\cdot G
s\rangle}{3\cdot2^{8}\pi^{4}}\Bigg\{\int_{\alpha_{min}}^{\alpha_{max}}\frac{d\alpha}{\alpha^{2}}\int_{\beta_{min}}^{1-\alpha}\frac{d\beta}{\beta^{2}}r\Big\{3m_{c}(\alpha+\beta-4\alpha\beta)r
+m_{s}\alpha\beta[12m_{c}^{2}-7(\alpha+\beta)m_{c}^{2}-5\alpha\beta s]\Big\}\nonumber\\
&&{}+\int_{\alpha_{min}}^{\alpha_{max}}d\alpha[m_{c}^{2}-\alpha(1-\alpha)s]\Big\{-\frac{3m_{c}}{\alpha(1-\alpha)}[m_{c}^{2}-\alpha(1-\alpha)s]+2m_{s}[5\alpha(1-\alpha)s-9m_{c}^{2}]\Big\}\Bigg\},\nonumber\\
\rho^{\langle\bar{s}s\rangle^{2}}&=&\frac{m_{c}\varrho\langle\bar{s}s\rangle^{2}}{3\cdot2^{4}\pi^{2}}\int_{\alpha_{min}}^{\alpha_{max}}d\alpha\Big\{-2m_{c}[m_{c}^{2}-\alpha(1-\alpha)s]-m_{s}[m_{c}^{2}-2\alpha(1-\alpha)s]\Big\},\nonumber\\
\rho^{\langle g^{3}G^{3}\rangle}&=&-\frac{\langle
g^{3}G^{3}\rangle}{3^{2}\cdot2^{14}\pi^{6}}\int_{\alpha_{min}}^{\alpha_{max}}\frac{d\alpha}{\alpha^{4}}\int_{\beta_{min}}^{1-\alpha}\frac{d\beta}{\beta^{4}}(1-\alpha-\beta)\kappa
\Big\{[(\alpha^{3}+\beta^{3})r+4(\alpha^{4}+\beta^{4})m_{c}^{2}\nonumber\\
&&{}+2m_{c}m_{s}(2\alpha^{2}+3\alpha\beta+2\beta^{2})(3\alpha^{2}-4\alpha\beta+3\beta^{2})]r+4m_{s}m_{c}^{3}(\alpha+\beta)(\alpha^{4}+\beta^{4})\Big\},\nonumber\\
\rho^{\langle\bar{s}s\rangle\langle g^{2}G^{2}\rangle}&=&\frac{m_{c}\langle\bar{s}s\rangle\langle
g^{2}G^{2}\rangle}{3^{2}\cdot2^{8}\pi^{4}}\Bigg\{\int_{\alpha_{min}}^{\alpha_{max}}\frac{d\alpha}{\alpha^{2}}\int_{\beta_{min}}^{1-\alpha}\frac{d\beta}{\beta^{2}}\Big\{-(2-\alpha-\beta)(\alpha^{3}+\beta^{3})m_{c}^{2}+3[\alpha^{2}(\beta-1)+\beta^{2}(\alpha-1)]r\nonumber\\
&+&[2(\alpha^{2}+\beta^{2})^{2}-(\alpha+\beta)^{3}-\alpha\beta(\alpha-\beta)^{2}-(\alpha^{3}+\beta^{3})]m_{s}m_{c}\Big\}
- m_{s}m_{c}\int_{\alpha_{min}}^{\alpha_{max}}d\alpha\frac{3\alpha^{2}-3\alpha+1}{\alpha(1-\alpha)}\Bigg\}
,\nonumber\\
\rho^{\langle\bar{s}s\rangle\langle g\bar{s}\sigma\cdot G s\rangle}&=&\frac{m_{c}(m_{c}+m_{s})\langle\bar{s}s\rangle\langle g\bar{s}\sigma\cdot G s\rangle}{3\cdot2^{5}\pi^{2}}\int_{\alpha_{min}}^{\alpha_{max}}d\alpha(6\alpha^{2}-6\alpha+1)
\nonumber
\end{eqnarray}
for the $P$-wave pseudoscalar-pseudoscalar case.
It is defined as $r=(\alpha+\beta)m_{c}^2-\alpha\beta s$ and $\kappa=1+\alpha-2\alpha^{2}+\beta+2\alpha\beta-2\beta^{2}$.
The integration limits are $\alpha_{min}=(1-\sqrt{1-4m_{c}^{2}/s})/2$,
$\alpha_{max}=(1+\sqrt{1-4m_{c}^{2}/s})/2$, and $\beta_{min}=\alpha
m_{c}^{2}/(s\alpha-m_{c}^{2})$.
For the four-quark condensate $\langle\bar{s}s\rangle^{2}$,
a general factorization $\langle\bar{s}s\bar{s}s\rangle=\varrho\langle\bar{s}s\rangle^{2}$ \cite{overview1,Narison}
has been used, where $\varrho$ is a constant that may be equal
to 1 or 2.

After equating the two expressions (\ref{Ph}) and (\ref{OPE}), assuming quark-hadron duality, and
making a Borel transform, the sum rule can be given by
\begin{eqnarray}\label{sumrule}
\lambda^{2}e^{-M_{H}^{2}/M^{2}}&=&\int_{(2m_{c}+2m_{s})^{2}}^{s_{0}}ds\rho e^{-s/M^{2}}.
\end{eqnarray}
Eliminating the hadronic coupling constant $\lambda$, one could
yield
\begin{eqnarray}\label{sum rule}
M_{H}^{2}&=&\int_{(2m_{c}+2m_{s})^{2}}^{s_{0}}ds\rho s
e^{-s/M^{2}}/
\int_{(2m_{c}+2m_{s})^{2}}^{s_{0}}ds\rho e^{-s/M^{2}}.
\end{eqnarray}

\section{Numerical analysis}\label{sec3}
Performing the numerical analysis of sum rule (\ref{sum rule}),
the $s$-quark and
the running charm quark masses
are chosen as updated values \cite{PDG}: $m_{s}=93_{-5}^{+11}~\mbox{MeV}$
and $m_{c}=1.27\pm0.02~\mbox{GeV}$, respectively.
Besides, other input parameters are taken as \cite{svzsum,overview3}:
$\langle\bar{q}q\rangle=-(0.24\pm0.01)^{3}~\mbox{GeV}^{3}$,
$m_{0}^{2}=0.8\pm0.1~\mbox{GeV}^{2}$,
$\langle\bar{s}s\rangle=m_{0}^{2}~\langle\bar{q}q\rangle$, $\langle
g\bar{s}\sigma\cdot G s\rangle=m_{0}^{2}~\langle\bar{s}s\rangle$,
$\langle
g^{2}G^{2}\rangle=0.88\pm0.25~\mbox{GeV}^{4}$, and $\langle
g^{3}G^{3}\rangle=0.58\pm0.18~\mbox{GeV}^{6}$.

Complying with the standard criterion
of sum rule analysis,
both the OPE convergence and the pole dominance would be considered
to find appropriate work windows for the threshold parameter $\sqrt{s_{0}}$ and the Borel
parameter $M^{2}$:
the lower bound of $M^{2}$ is gained by analyzing the OPE
convergence, and
the upper one is obtained by
viewing that the pole contribution should be larger
than QCD continuum contribution.
At the same time,
the threshold
$\sqrt{s_{0}}$ characterizes the beginning
of continuum state and is empirically about
$400\sim600~\mbox{MeV}$ above the extracted
$M_{H}$.

Taking the analysis of $P$-wave
scalar-scalar case as an example,
the input parameters would be first kept at their central values.
To obtain the lower bound of $M^{2}$, the OPE
convergence is shown in FIG. 1 by comparing the relative contributions of different
condensates from sum rule (\ref{sumrule}) for $\sqrt{s_{0}}=5.2~\mbox{GeV}$.
In numerical, the relative
perturbative contribution begins to play a
dominant role in the OPE side at $M^{2}=3.0~\mbox{GeV}^{2}$, which is increasing
with the Borel parameter $M^{2}$. Thereby,
the perturbative part could dominate in OPE
comparing with other higher dimensional
condensate contributions while taking $M^{2}\geq3.0~\mbox{GeV}^{2}$.
On the other hand, the upper bound of $M^{2}$ is gained by
considering the pole dominance phenomenologically.
In FIG. 2, the comparison
between pole and continuum contributions from sum rule (\ref{sumrule})
is shown for $\sqrt{s_{0}}=5.2~\mbox{GeV}$.
The relative pole contribution
is approximate to $50\%$ at $M^{2}=3.5~\mbox{GeV}^{2}$ and descending with $M^{2}$.
Hence, the pole contribution dominance could be satisfied
when $M^{2}\leq3.5~\mbox{GeV}^{2}$.
Consequently, the Borel window of $M^{2}$ is fixed on
$3.0\sim3.5~\mbox{GeV}^{2}$ for $\sqrt{s_0}=5.2~\mbox{GeV}$.
In the similar analysis,
the proper range of $M^{2}$ is gained as $3.0\sim3.4~\mbox{GeV}^{2}$ for $\sqrt{s_0}=5.1~\mbox{GeV}$, and
 $3.0\sim3.7~\mbox{GeV}^{2}$ for $\sqrt{s_0}=5.3~\mbox{GeV}$.
In the chosen work windows,
it is expected that the two sides of QCD sum rules have a good
overlap and information on the resonance can be safely
extracted.
The mass $M_{H}$ of $P$-wave scalar-scalar $[cs][\bar{c}\bar{s}]$ state
is shown as a function of $M^{2}$ in FIG. 3,
and it is
computed to be
$4.60\pm0.10~\mbox{GeV}$ in work windows.
Next, varying the quark masses and condensates, one could arrive at
$4.60\pm0.10^{+0.03}_{-0.04}~\mbox{GeV}$ (the first error resulted from
the uncertainty due
to variation of $s_{0}$ and
$M^{2}$, and the second error rooting in the variation of QCD parameters) or briefly $4.60^{+0.13}_{-0.14}~\mbox{GeV}$.
At last, taking into account the variation of four-quark condensate
factorization factor $\varrho$ from 1 to 2,
one could get the final mass value $4.60^{+0.13}_{-0.19}~\mbox{GeV}$ for the $P$-wave scalar-scalar case,
which is in good agreement with the experimental data of $Y(4626)$
and could support its $P$-wave scalar-scalar $[cs][\bar{c}\bar{s}]$ tetraquark explanation.

\begin{figure}[htb!]
\centerline{\epsfysize=7.18truecm\epsfbox{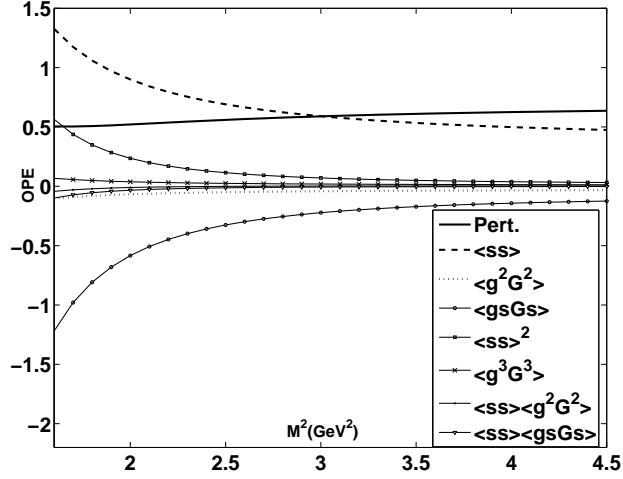}}
\caption{The OPE convergence for the $P$-wave scalar-scalar case is shown by comparing the relative contributions of
perturbative, two-quark condensate $\langle\bar{s}s\rangle$, two-gluon condensate $\langle
g^{2}G^{2}\rangle$, mixed condensate $\langle
g\bar{s}\sigma\cdot G s\rangle$, four-quark condensate $\langle\bar{s}s\rangle^{2}$, three-gluon
condensate $\langle g^{3}G^{3}\rangle$,
$\langle\bar{s}s\rangle\langle g^{2}G^{2}\rangle$, and
$\langle\bar{s}s\rangle\langle g\bar{s}\sigma\cdot G s\rangle$
from sum rule (\ref{sumrule})
for $\sqrt{s_{0}}=5.2~\mbox{GeV}$.}
\end{figure}

\begin{figure}[htb!]
\centerline{\epsfysize=7.18truecm\epsfbox{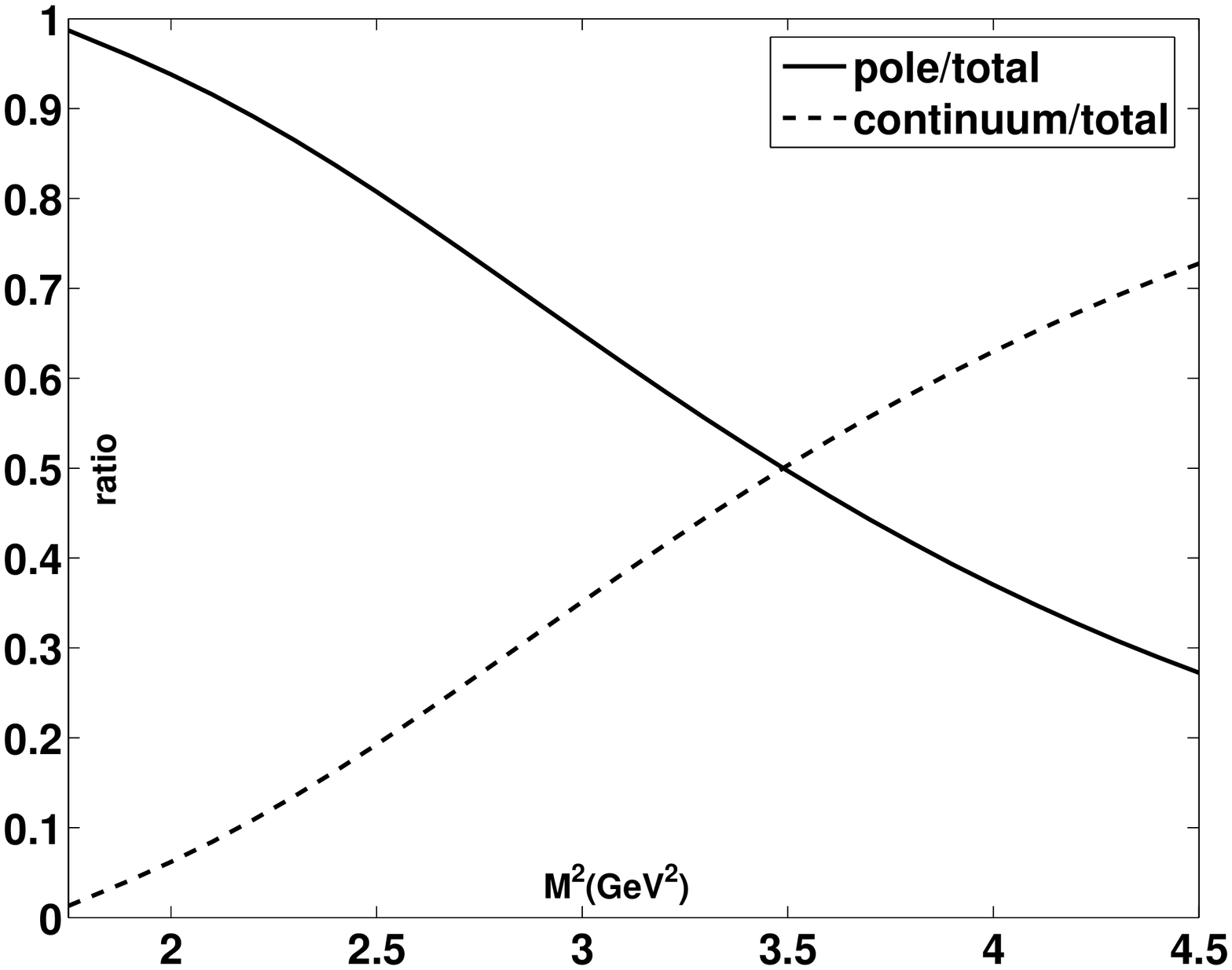}}
\caption{The phenomenological contribution in sum rule
(\ref{sumrule}) for $\sqrt{s_{0}}=5.2~\mbox{GeV}$ for the $P$-wave scalar-scalar case.
The solid line is the relative pole contribution (the pole
contribution divided by the total, pole plus continuum contribution)
as a function of $M^2$ and the dashed line is the relative continuum
contribution.}
\end{figure}

\begin{figure}[htb!]
\centerline{\epsfysize=7.18truecm
\epsfbox{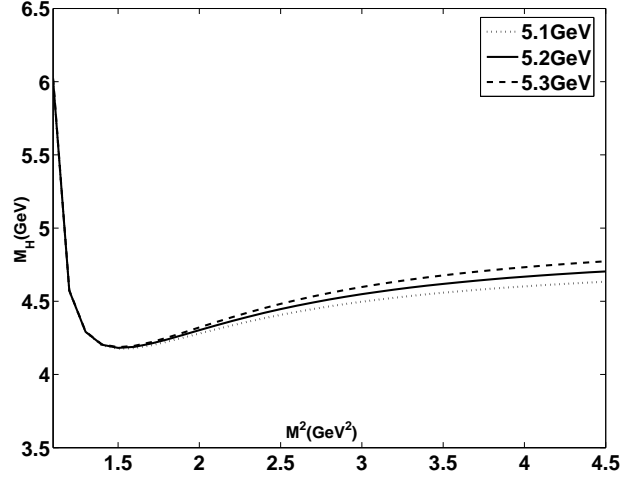}}\caption{The
dependence on $M^2$ for the mass $M_{H}$ of $P$-wave scalar-scalar
$[cs][\bar{c}\bar{s}]$ from sum rule (\ref{sum rule}) is shown.
The ranges of $M^{2}$ are $3.0\sim3.4~\mbox{GeV}^{2}$ for
$\sqrt{s_0}=5.1~\mbox{GeV}$, $3.0\sim3.5~\mbox{GeV}^{2}$
for $\sqrt{s_0}=5.2~\mbox{GeV}$, and
$3.0\sim3.7~\mbox{GeV}^{2}$ for $\sqrt{s_0}=5.3~\mbox{GeV}$, respectively.}
\end{figure}

For the $P$-wave pseudoscalar-pseudoscalar case, FIG. 4 shows
the relative contributions of
different condensates from sum rule (\ref{sumrule})
for $\sqrt{s_{0}}=5.2~\mbox{GeV}$.
Note that the two-quark condensate $\langle\bar{s}s\rangle$
plays an important role on OPE here and thus the lower bound of
$M^{2}$ has to be taken a very high value to meet the convergence condition.
Whereas, its phenomenological contribution in sum rule
(\ref{sumrule}) is shown in FIG. 5 for $\sqrt{s_{0}}=5.2~\mbox{GeV}$,
and the Borel parameter should be taken $M^{2}\leq2.5~\mbox{GeV}^{2}$
to fulfill the pole contribution dominance.
For this case, one could note that it is difficult to find reasonable work windows
satisfying both good OPE convergence and pole dominance, and graphically
the mass $M_{H}$'s dependence on $M^2$ in FIG. 6
is rather unstable.
Accordingly, it is not advisable to continue extracting a mass value.
In a sideward way, it is consistent with
the statement that diquarks are preferably formed into
$0^{+}$ good diquarks.

\begin{figure}[htb!]
\centerline{\epsfysize=7.18truecm\epsfbox{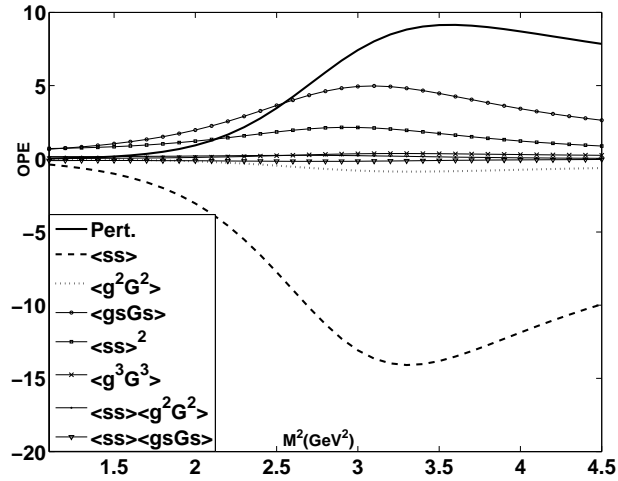}}
\caption{The OPE convergence for the $P$-wave pseudoscalar-pseudoscalar case is shown by comparing the relative contributions of
perturbative, two-quark condensate $\langle\bar{s}s\rangle$, two-gluon condensate $\langle
g^{2}G^{2}\rangle$, mixed condensate $\langle
g\bar{s}\sigma\cdot G s\rangle$, four-quark condensate $\langle\bar{s}s\rangle^{2}$, three-gluon
condensate $\langle g^{3}G^{3}\rangle$,
$\langle\bar{s}s\rangle\langle g^{2}G^{2}\rangle$, and
$\langle\bar{s}s\rangle\langle g\bar{s}\sigma\cdot G s\rangle$
from sum rule (\ref{sumrule})
for $\sqrt{s_{0}}=5.2~\mbox{GeV}$.}
\end{figure}

\begin{figure}[htb!]
\centerline{\epsfysize=7.18truecm\epsfbox{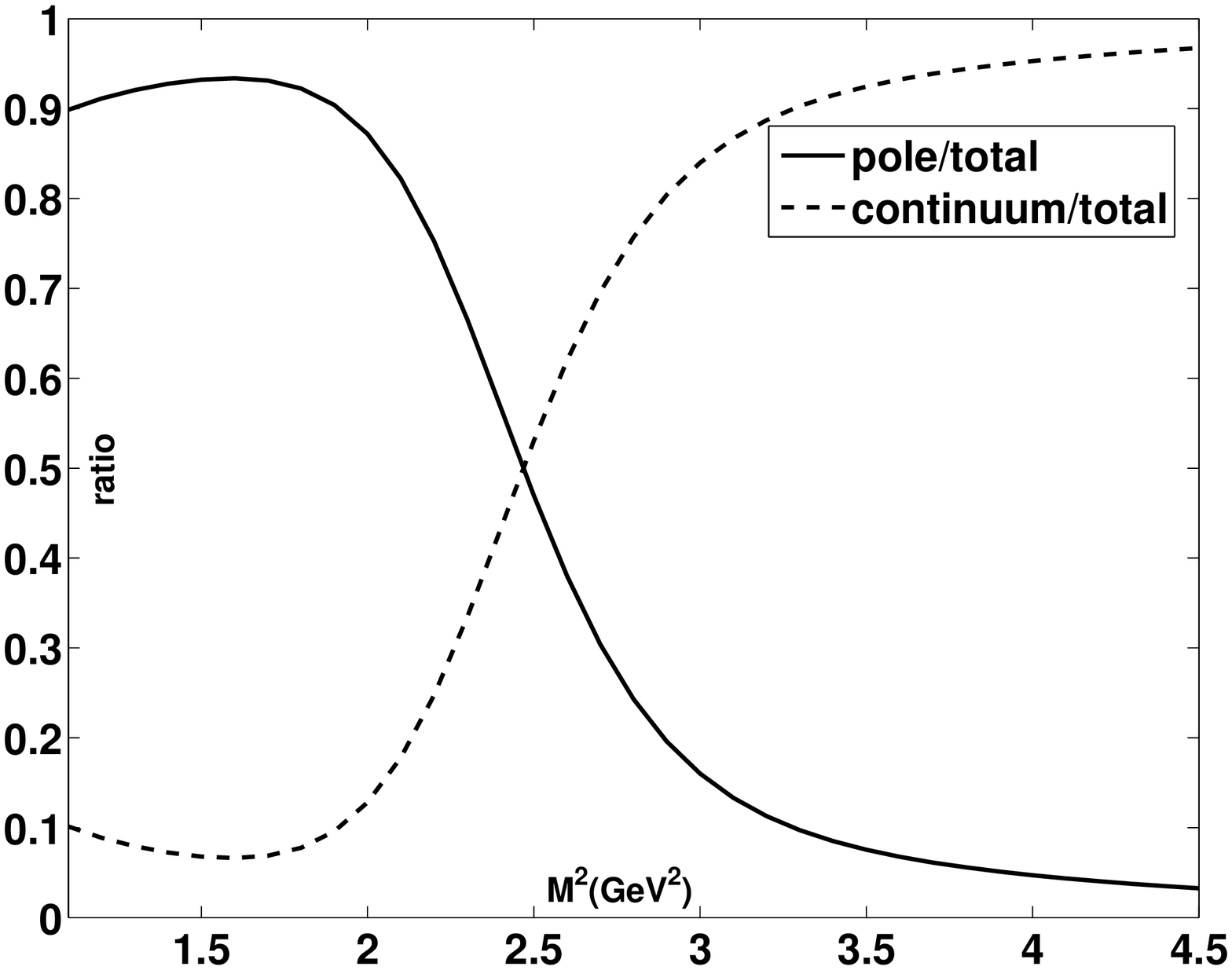}}
\caption{The phenomenological contribution in sum rule
(\ref{sumrule}) for $\sqrt{s_{0}}=5.2~\mbox{GeV}$ for the $P$-wave pseudoscalar-pseudoscalar case.
The solid line is the relative pole contribution (the pole
contribution divided by the total, pole plus continuum contribution)
as a function of $M^2$ and the dashed line is the relative continuum
contribution.}
\end{figure}

\begin{figure}[htb!]
\centerline{\epsfysize=7.18truecm
\epsfbox{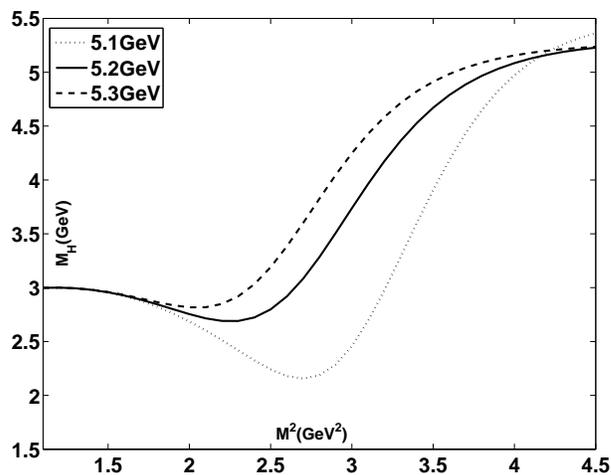}}\caption{The
dependence on $M^2$ for the mass $M_{H}$ of $P$-wave
pseudoscalar-pseudoscalar tetraquark state from sum rule (\ref{sum rule}) is shown.}
\end{figure}

\section{Summary and outlook}\label{sec4}
Stimulated by the first observation of a vector
charmoniumlike state $Y(4626)$ decaying to a $D_{s}^{+}D_{s1}(2536)^{-}$ pair,
we have calculated the
mass of $P$-wave $[cs][\bar{c}\bar{s}]$ tetraquark state
in QCD sum rules.
For
the $P$-wave scalar-scalar case, the final result $4.60^{+0.13}_{-0.19}~\mbox{GeV}$
is well compatible with the experimental data
$4625.9_{-6.0}^{+6.2}
\pm0.4~\mbox{MeV}$ of $Y(4626)$, which favors the explanation of $Y(4626)$
as a $P$-wave scalar-scalar $[cs][\bar{c}\bar{s}]$ tetraquark state.
It is of difficulty to find proper work windows
to achieve a mass value
for the $P$-wave pseudoscalar-pseudoscalar case,
in a sideward way, which is coincident with the
picture of $Y(4626)$
as a $P$-wave scalar-scalar $[cs][\bar{c}\bar{s}]$ tetraquark state.

In the present article,
we have devoted to calculating the mass of $P$-wave $[cs][\bar{c}\bar{s}]$ state from
two-point QCD sum rules, which could provide some evidence for
the newly observed $Y(4626)$ as a $P$-wave scalar-scalar
$[cs][\bar{c}\bar{s}]$ tetraquark state.
To finalize the inner structure of $Y(4626)$,
undoubtedly it needs further experimental observations
and continually theoretical studies. For instance,
one could take into account
studying the width of the state,
which is definitely important and
could be obtained by
employing three-point QCD sum rules. But then, one can expect that there
many Feynman diagrams should be considered
particularly owing to the derivative operator of interpolating current,
which could be researched in some subsequent work after having completed
enormous calculations.
Anyhow, one can expect that future experimental
and theoretical efforts may shed more light on the nature of
$Y(4626)$.

\begin{acknowledgments}
This work was supported by the National
Natural Science Foundation of China under Contract
Nos. 11475258 and 11675263, and by the project for excellent youth talents in
NUDT.
\end{acknowledgments}


\end{document}